\documentclass[runningheads]{llncs}
\usepackage[T1]{fontenc}
\usepackage{booktabs}
\usepackage{multirow}
\usepackage{amsmath,amssymb,amsfonts}
\usepackage{array}
\usepackage[normalem]{ulem}
\usepackage{graphicx,verbatim}
\begin{document}
\title{Tau Anomaly Detection in PET Imaging via Bilateral-Guided Deterministic Diffusion Model}
%

\author{%
  Lujia Zhong\textsuperscript{1,2}, Shuo Huang\textsuperscript{1,3}, Jiaxin Yue\textsuperscript{1,2},\\
  Jianwei Zhang\textsuperscript{1,2}, Zhiwei Deng\textsuperscript{1,2}, Wenhao Chi\textsuperscript{1,2}, Yonggang Shi\textsuperscript{1,2,3}
}
\institute{
\textsuperscript{1} Stevens Neuroimaging and Informatics Institute, Keck School of Medicine, University of Southern California \\
  \textsuperscript{2} Ming Hsieh Department of Electrical and Computer Engineering, Viterbi School of
Engineering, University of Southern California \\
  \textsuperscript{3} Alfred E. Mann Department of Biomedical Engineering, Viterbi School of
Engineering, University of Southern California \\}
    
\maketitle              
\begin{abstract}
The emergence of tau PET imaging over the last decade has enabled Alzheimer's disease (AD) researchers to examine tau pathology in vivo and more effectively characterize the disease trajectories of AD. Current tau PET analysis methods, however, typically perform inferences on large cortical ROIs and are limited in the detection of localized tau pathology that varies across subjects. In this work, we propose a novel bilateral-guided deterministic diffusion sampling method to perform anomaly detection from tau PET imaging data. By including individualized brain structure and cognitively normal (CN) template conditions, our model computes a voxel-level anomaly map based on the deterministically sampled pseudo-healthy reconstruction. We train our model on ADNI CN subjects (n=380) and evaluate anomaly localization performance on the left MCI/AD subjects (n=154) and the preclinical subjects of the A4 clinical trial (n=447). We further train a CNN classifier on the derived 3D anomaly maps from ADNI, including CN and MCI/AD, to classify subjects into two groups and test classification performance on A4. We demonstrate that our method outperforms baselines in anomaly localization. Additionally, we show that our method can successfully group preclinical subjects with significantly different cognitive functions, highlighting the potential of our approach for application in preclinical screening tests. The code will be publicly available.

\keywords{Tau Anomaly  \and Diffusion Model \and Probability Flow ODE.}

\end{abstract}
\section{Introduction}

There is an explosion of interest in tau pathology in AD studies because of its strong correlation with clinical symptoms \cite{nelson2012correlation} and the increasing availability of tau PET imaging \cite{SCHOLL2016971}. Previous tau PET studies typically rely on canonical cortical parcellation to perform ROI-level inferences of tau pathology \cite{SCHOLL2016971}, and there is a lack of robust tools for the automated detection of localized tau pathology that varies across subjects.

For the localized detection of anomalies in tau PET imaging data, Z-score-based normative analysis can be applied in a common space 
\cite{akamatsu2019voxel}, but the varied cortical folding and resulting misalignment will no doubt compromise the accurate detection of tau pathology. Recent works utilizing VAEs \cite{choi2019deep,hassanaly2024evaluation}, flow-based models \cite{xiong2023pet}, and GANs \cite{han2021madgan,siddiquee2019learning,baydargil2021anomaly,yang2023learning} hold the promise of more accurate individualized reconstruction and anomaly localization. Denoising diffusion probabilistic models (DDPM) \cite{ho2020denoising} recently have been applied to medical anomaly detection because of their superior generation ability and controllability 
\cite{wolleb2022diffusion,kascenas2023role,behrendt2024patched}. The application to tau pathology detection, however, is limited by suboptimal reconstruction of pseudo-healthy data.

In this work, we propose a bilateral-guided deterministic sampling method for the diffusion model that extends the DDPM with individualized brain structure constraint and CN template condition during inference. We also mathematically derive a deterministic sampling method by approximating our bilateral-guided ancestral sampling to probability flow ordinary differential equation (ODE) under the stochastic differential equation (SDE) framework \cite{song2020score}. Our deterministic sampling method can be directly applied to a pre-trained diffusion model without the need for fine-tuning. Compared with baseline models of GAN, VAE, DDPM \cite{ho2020denoising}, and AnoDiff \cite{wolleb2022diffusion}, our methods show superiority in anomaly localization and preclinical subject grouping.

\section{Preliminary}

\subsection{Stochastic Differential Equations (SDE)}

Yang Song, et al. \cite{song2020score} unify the score-based diffusion models and denoising diffusion probabilistic models under the framework of stochastic differential equations by modeling the forward and backward processes of diffusion models as SDE and reverse SDE. The general forward and reverse SDE forms are defined as:
\begin{align*}
\text{(forward)} \quad d\mathbf{x} =& \mathbf{f}(\mathbf{x}, t) dt + \mathbf{G}(\mathbf{x}, t) d\mathbf{w} \\
\text{(reverse)} \quad d\mathbf{x} =& \{ \mathbf{f}(\mathbf{x}, t) - \nabla \cdot [\mathbf{G}(\mathbf{x},t)\mathbf{G}(\mathbf{x},t)^T] \nonumber \\
&- \mathbf{G}(\mathbf{x},t)\mathbf{G}(\mathbf{x},t)^T \nabla_{\mathbf{x}} \log p_t(\mathbf{x}) \} dt + \mathbf{G}(\mathbf{x},t) d\mathbf{\bar{w}}
\end{align*}

where \( \mathbf{f}(\mathbf{x},t) \) and \( \mathbf{G}(\mathbf{x},t) \) are the drift and the diffusion coefficients, respectively; \( d\mathbf{w} \) and \( d\mathbf{\bar{w}} \) represent forward and reverse standard Wiener processes (Brownian motions), respectively. The \( \nabla_{\mathbf{x}} \log p_t(\mathbf{x}) \) is the gradient of the log probability density (score function) with respect to \( \mathbf{x} \) at time t, which is what neural network learns to denoise and reconstruct the data.

\subsection{Probability Flow Ordinary Differential Equations (ODE)}

Under the framework of SDE, one can derive the probability flow ODE of the diffusion process by reparameterizing the Kolmogorove's forward equation (Fokker-Planck equation):
\begin{align*}
\frac{\partial p_t(\mathbf{x})}{\partial t} &= - \sum_{i=1}^{d}{\frac{\partial}{\partial x_i} \left[ f_i(\mathbf{x},t) p_t(\mathbf{x}) \right] } \nonumber \\
&+ \frac{1}{2} \sum_{i=1}^{d}\sum_{j=1}^{d} \frac{\partial^2}{\partial x_i \partial x_j} \left[ \sum_{k=1}^{d} G_{ik}(\mathbf{x},t)G_{jk}(\mathbf{x},t) p_t(\mathbf{x}) \right]
\end{align*}

to
\begin{align}
\label{reparameterized_kolmogorove}
\frac{\partial p_t(\mathbf{x})}{\partial t} = - \sum_{i=1}^{d}{\frac{\partial}{\partial x_i} \left[ \widetilde{f}_i(\mathbf{x},t) p_t(\mathbf{x}) \right] }
\end{align}

where the \( \widetilde{\mathbf{f}}(\mathbf{x},t) \) is defined as:
\begin{align*}
\widetilde{\mathbf{f}}(\mathbf{x},t) := \mathbf{f}(\mathbf{x},t) - \frac{1}{2} \nabla \cdot [\mathbf{G}(\mathbf{x},t)\mathbf{G}(\mathbf{x},t)^T] \nonumber
- \frac{1}{2}\mathbf{G}(\mathbf{x},t)\mathbf{G}(\mathbf{x},t)^T \nabla_{\mathbf{x}} \log p_t(\mathbf{x})
\end{align*}

The reparameterized equation in \eqref{reparameterized_kolmogorove} is equivalent to Kolmogorove's forward equation of the following ODE:
\begin{align*}
d\mathbf{x} = \mathbf{\widetilde{f}}(\mathbf{x}, t) dt
\end{align*}

Therefore, we end up with the following probability flow ODE form for the reverse SDE:
\begin{align*}
d\mathbf{x} = \{ \mathbf{f}(\mathbf{x}, t) - \frac{1}{2} \nabla \cdot [\mathbf{G}(\mathbf{x},t)\mathbf{G}(\mathbf{x},t)^T] \nonumber - \frac{1}{2} \mathbf{G}(\mathbf{x},t)\mathbf{G}(\mathbf{x},t)^T \nabla_{\mathbf{x}} \log p_t(\mathbf{x}) \} dt
\end{align*}

\section{Methodology}

\begin{figure}
\centering
\includegraphics[width=1\columnwidth]{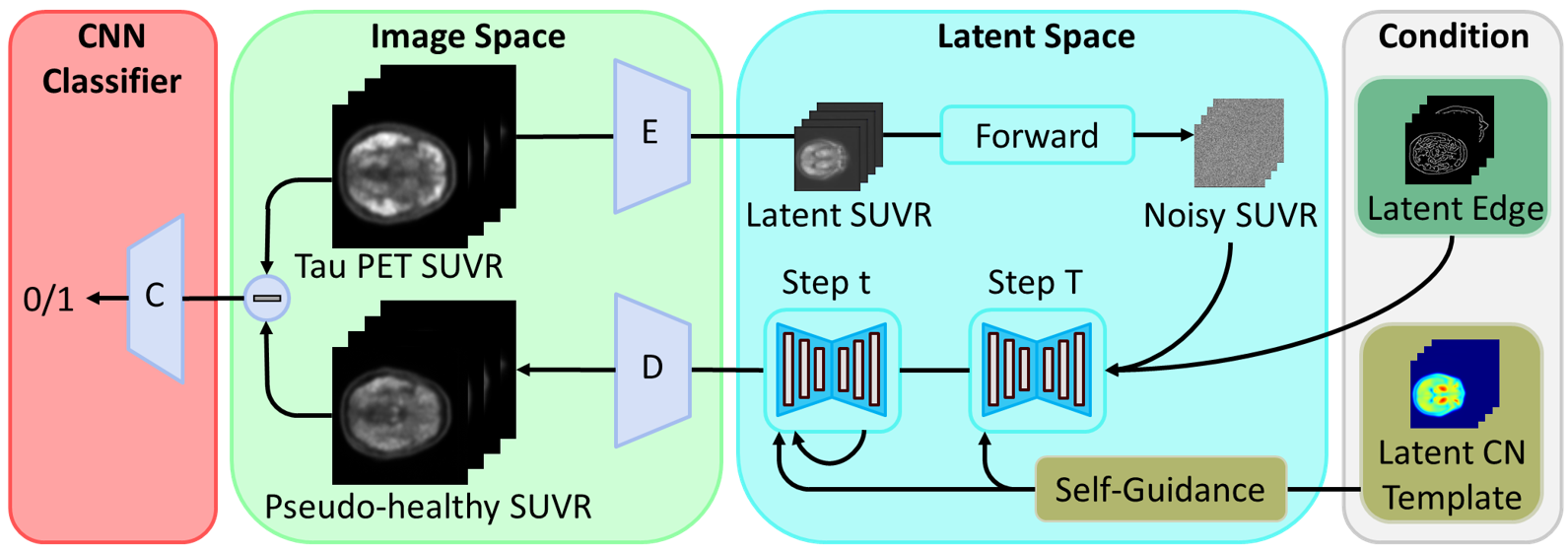}
\caption{The inference workflow with our bilateral-guided diffusion model.}
\label{workflow}
\end{figure}

In this work, we propose a bilateral-guided diffusion model, utilizing both the 3D template derived from averaging CN subjects' SUVR data and an edge map for individualized brain structure constraint to guide accurate pseudo-healthy reconstruction. Because of the stochastic nature of diffusion models, one may have to run reconstructions multiple times and calculate the average as a final reconstruction for better results. This is effective but unnecessarily time-consuming. To tackle this, we derive two versions of deterministic sampling for DDPM by approximating the reverse diffusion process of DDPM to the reverse SDE. The overall workflow is demonstrated in Figure~\ref{workflow}.

\subsection{3D Latent Conditioned Diffusion Model}

Diffusion models are easy to train compared to GANs and can provide high-quality reconstructions. However, it can be time-consuming during inference because of multiple denoising steps, which is even worse when dealing with 3D data. Hence, inspired by \cite{rombach2022high}, we first train a 3D autoencoder to compress the data in a smaller latent space. Then, we train a 3D diffusion model conditioned on a 3D edge map derived from the latent tau PET data in this latent space. The edge map is concatenated with the data channel-wisely to force brain structure-aligned reconstruction.





\subsection{Bilateral Guidance}

Similar to previous works, our model's training process only involves CN subjects. During inference, the trained conditional DDPM will remove anomalies (high tau) from the given data while maintaining an individualized brain structure. However, when the anomaly data is located far away from the CN subjects distribution in the high-dimensional space, diffusion-based methods still cannot remove all tau anomalies, although with multiple denoising steps compared to traditional GANs and VAEs. Therefore, we introduce additional intensity guidance during sampling, inspired by \cite{epstein2023diffusion}, to generate more realistic pseudo-healthy data. This intensity guidance, combined with brain structure guidance, is called bilateral guidance, which is similar to the terminology in bilateral filtering \cite{tomasi1998bilateral}.

Specifically, we construct a CN template $\mathcal{T}$ by averaging ADNI CN data and use it to guide the latent diffusion model sampling trajectory during inference, according to the general sampling formula:
\begin{align*}
    \hat{\boldsymbol{\epsilon}}_t = (1 + s) \boldsymbol{\epsilon}_{\theta}(\mathbf{x}_t; t, y) - s \boldsymbol{\epsilon}_{\theta}(\mathbf{x}_t; t, \emptyset) + \nu \sigma_t \nabla_{\mathbf{x}_t} g(\mathbf{x}_t; t, y)
\end{align*}

where the $g$ is an energy function and is formulated as below following \cite{epstein2023diffusion}:
\begin{align*}
    &g(\mathbf{x}_t; t, y) = \left\| \text{appearance}_{t,\mathbf{x}_t} - \text{appearance}_{t,y} \right\|_1\\
    &\text{appearance}_{t} = \frac{\sum_{h,w,d} \text{shape} \odot \Psi_t}{\sum_{h,w,d} \text{shape}}
\end{align*}

where $t$ represents the time steps during sampling; the shape is a binary segmentation map indicating the brain area in our case; $\Psi_t$ is chosen to be the reconstructed $X_0$ in each time step instead of activation maps as in \cite{epstein2023diffusion}. Because we are using \textit{v-prediction} \cite{salimans2022progressive} without classifier-free guidance, we finally have:
\begin{align*}
    \hat{\boldsymbol{\epsilon}}_t = \frac{1}{\sqrt{\bar{\alpha}_t}} \mathbf{v}_{\theta}(\mathbf{x}_t; t) + \sqrt{\frac{1}{\bar{\alpha}_t} - 1} (\sqrt{\bar{\alpha}_t} \mathbf{x}_t - \sqrt{1 - \bar{\alpha}_t} \mathbf{v}_{\theta}(\mathbf{x}_t; t)) \nonumber + \nu \sigma_t \nabla_{\mathbf{x}_t} g(\mathbf{x}_t; t, \mathcal{T})
\end{align*}

\subsection{Deterministic Sampling from Probability Flow ODE}

To avoid averaging multiple runs of diffusion models for better efficiency, we derive a deterministic sampling method of DDPM. Specifically, we approximate DDPM sampling (ancestral sampling) to reverse SDE form in two ways:

(1) Consider the diffusion coefficients of DDPM as $\mathbf{G}(\mathbf{x}, t)$ of SDE:

\begin{align*}
    \mathbf{G}(\mathbf{x}, t) = G_{1}(t) = \sqrt{\frac{1 - \bar{\alpha}_{t-1}}{1 - \bar{\alpha}_{t}} (1 - \alpha_t)}
\end{align*}

Then we have DDPM ancestral sampling:
\begin{align*}
    \mathbf{x}_{t-1} &= \mathbf{x}_t - \mathbf{f}(\mathbf{x}, t) + \frac{1 - \bar{\alpha}_{t}}{1 - \bar{\alpha}_{t-1}} \frac{1}{\sqrt{\alpha_t}} G_{1}(t)^2 \mathbf{S}_{\theta} + G_{1}(t) \boldsymbol{\epsilon} \nonumber \\
    &\approx \mathbf{x}_t - \mathbf{f}(\mathbf{x}, t) + G_{1}(t)^2 \mathbf{S}_{\theta} + G_{1}(t) \boldsymbol{\epsilon}
\end{align*}

which is equivalent to the probability flow ODE as below:
\begin{equation}
    \mathbf{x}_{t-1} = \mathbf{x}_t - \mathbf{f}(\mathbf{x}, t) + \frac{1}{2} G_{1}(t)^2 \mathbf{S}_{\theta}
    \label{ode1}
\end{equation}

(2) Consider the drift coefficients of DDPM as $\mathbf{G}(\mathbf{x}, t)\mathbf{G}(\mathbf{x}, t)^T$ of SDE:
\begin{align*}
    \mathbf{G}(\mathbf{x}, t)\mathbf{G}(\mathbf{x}, t)^T = G_{2}(t)^2 = \frac{1 - \alpha_t}{\sqrt{\alpha_t}}
\end{align*}

Similarly, we end up with sampling formula as:
\begin{align*}
    \mathbf{x}_{t-1} &= \mathbf{x}_t - \mathbf{f}(\mathbf{x}, t) + G_{2}(t)^2 \mathbf{S}_{\theta} \nonumber + (G_{2}(t) + \sqrt{1 - \alpha_t} (\sqrt{\frac{1 - \bar{\alpha}_{t-1}}{1 - \bar{\alpha}_{t}}} - \sqrt{\frac{1}{\sqrt{\alpha_t}}}) ) \boldsymbol{\epsilon} \nonumber \\
    &\approx \mathbf{x}_t - \mathbf{f}(\mathbf{x}, t) + G_{2}(t)^2 \mathbf{S}_{\theta} + G_{2}(t) \boldsymbol{\epsilon}
\end{align*} 

which is equivalent to the probability flow ODE:
\begin{equation}
    \mathbf{x}_{t-1} = \mathbf{x}_t - \mathbf{f}(\mathbf{x}, t) + \frac{1}{2} G_{2}(t)^2 \mathbf{S}_{\theta}
    \label{ode2}
\end{equation}

With the above probability flow ODEs in \eqref{ode1} and \eqref{ode2}, we can sample pseudo-healthy reconstruction deterministically. Note that the formula is a little bit different in practice because we estimate velocity $\mathbf{v}$ instead of noise $\boldsymbol{\epsilon}$ or score function $\mathbf{S}$, so the final formula involves $\mathbf{v}_{\theta}$ instead of $\mathbf{S}_{\theta}$.

\subsection{Classification and Anomaly Score}

With bilateral guidance and deterministic sampling, we end up with a pseudo-healthy reconstruction for input data. By subtracting the pseudo-healthy reconstruction from the input data, we acquire the anomaly map. After collecting all such anomaly maps in the training dataset of ADNI, including healthy and unhealthy subjects, we train a 3D CNN classifier on these anomaly maps to classify the data into positive/negative groups. Then, we apply it to the A4 dataset, which contains preclinical subjects without diagnosis labels.

We also derive a subject-level anomaly score by combining two factors: 1) $m_\text{SUVR}$: mean SUVR values within the brain area and 2) $p_\text{cls}$: classifier-derived probability. These two factors collectively give an anomaly score for each subject with geometric mean as below because they have different intensity scales:

\begin{equation*}
    \text{Anomaly Score} = \sqrt{m_\text{SUVR} \cdot p_\text{cls}}
\end{equation*}

\section{Experiments}

\subsection{Datasets}

We train our conditioned diffusion models on CN subjects (n=380) from the Alzheimer's Disease Neuroimaging Initiative (ADNI) \cite{Mueller05} and apply the trained model to the MCI/AD subjects (n=154) from ADNI and preclinical subjects (n=447) from the Anti-Amyloid Treatment in Asymptomatic Alzheimer's Disease study (A4) \cite{Sperling20}.

ADNI subjects with tau PET scans were screened for inclusion in our study. We used a tau positivity criterion defined as follows: 95 percentile of tau SUVR $>$1.4 \cite{weigand2022s}. In total, 380 tau-negative CN subjects, 111 tau-positive MCI, and 43 tau-positive AD subjects were included in our work. The A4 cohort includes 447 subjects with tau PET scans. All A4 subjects have a global clinical dementia rating (CDR) score of zero. The A4 data are exclusively used for final evaluation without being used for model training or validation in any form.

All tau PET scans used in this work are acquired with the AV1451/Flortaucipir (18F) tracer and were nonlinearly registered to the same space with Elastix \cite{klein2009elastix}. We obtain the standardized uptake value ratio (SUVR) by dividing data by the mean value of the inferior cerebellum. After preprocessing, we resize all data to 160×160×160 as our 3D autoencoder inputs.

To evaluate the classification results, we use the following cognitive assessments of the A4 study: COGDIGIT, COGFCSR16, and COGLOGIC \cite{psychometric2019,jaeger2018digit}; and we utilize ADNI-Mem, ADNI-EF, ADNI-Lan, and ADNI-VS \cite{choi2020development} for ADNI. For more details about these cognitive tests, we recommend referring to the A4 \cite{Sperling20} and ADNI studies \cite{Mueller05}.

\subsection{Implementation Details}

Both the 3D Autoencoder model and 3D conditioned latent diffusion model are trained exclusively on ADNI CN subjects, where we choose a training to validation data split ratio of 9:1. The 3D CNN classifier is trained on the 3D anomaly maps of ADNI data, involving both CN subjects as a negative group and MCI/AD subjects as a positive group. The Autoencoder is trained with reconstruction loss, which compresses the 3D tau PET data from image space with the shape of (160, 160, 160) to a latent space (40, 40, 40) and then reconstructs it back to the image space. To train the 3D diffusion model, the condition of edge maps is calculated in the latent space. The autoencoder implementation follows \cite{rombach2022high}, and the CNN classifier contains 3 CNN layers and 2 MLP layers, where both are trained with a batch size of 1 and Adam optimizer with betas of (0.5, 0.9) and a learning rate of 2e-4. The conditioned latent diffusion model with a similar architecture to \cite{ho2020denoising} with a batch size of 4 and AdamW optimizer with betas of (0.9, 0.999) and a learning rate of 2e-4. Both models are trained for 100,000 iterations. Following \cite{salimans2022progressive}, we use \textit{v-prediction} for diffusion model training. The training costs about 12 hours for the Autoencoder and 40 hours for the diffusion model on a single NVIDIA RTX A5000 GPU.

\begin{figure}
    \centering
    \includegraphics[width=1\linewidth]{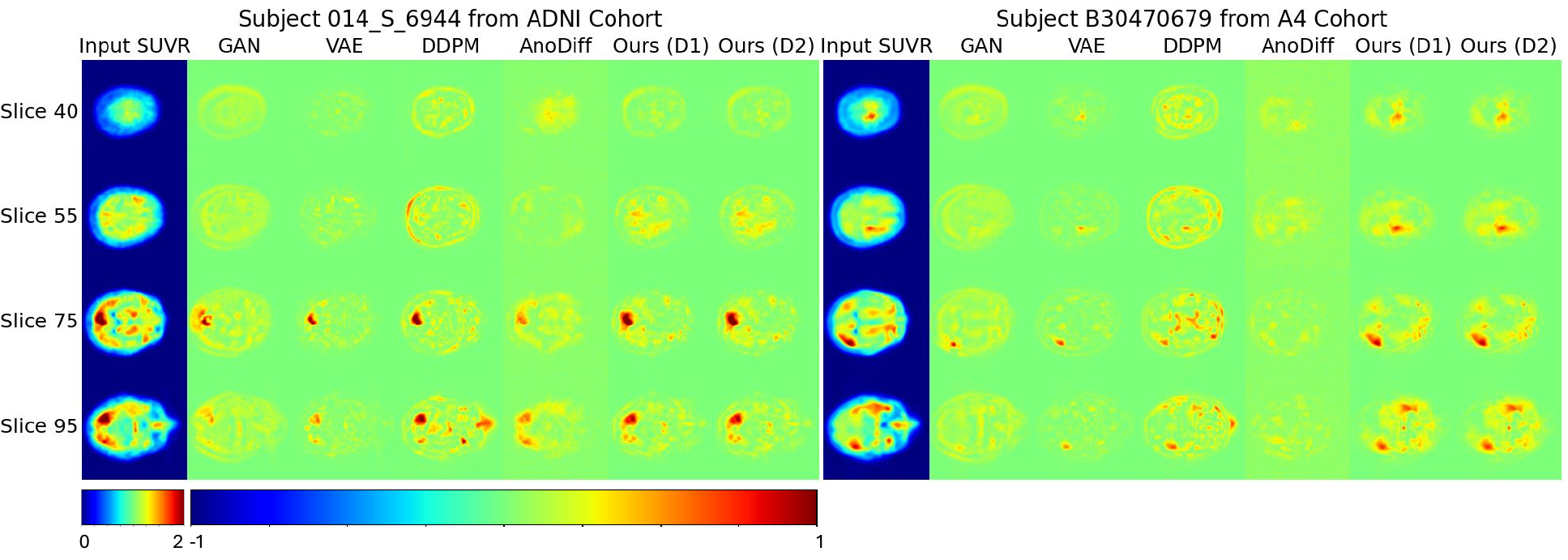}
    \caption{Anomaly localization comparisons on ADNI and A4 subjects.}
    \label{fig:anomaly}
\end{figure}

\subsection{Results}

We first demonstrate the anomaly localization performance of our method on ADNI tau-positive subjects and preclinical A4 subjects. As shown in Figure~\ref{fig:anomaly}, we present two representative subjects \textit{014\_S\_6944} from ADNI and \textit{B30470679} from A4, where the plots are presented by the difference between the generated pseudo-healthy version of input data and the input SUVR for each method. Both deterministic versions of our methods, where D1 for Eq.~\eqref{ode1} and D2 for Eq.~\eqref{ode2}, produce anomaly patterns that are more closely aligned with the input SUVR in cortical areas compared to all baselines. AnoDiff shows reasonable localization ability only for the ADNI subject, which has more severe tau anomalies. The vanilla DDPM exhibits less alignment of brain structures for the generated pseudo-healthy data compared to the input SUVR, leading to false positives around the brain. Both GAN and VAE also show suboptimal anomaly localization compared to our methods.

\begin{table}[h!]
\centering
\caption{Pearson correlation between anomaly scores and SUVR across methods and brain lobes on A4}
\label{tab:method_corr}
\begin{tabular}{lcccccccc}
\toprule
& \multicolumn{4}{c}{ADNI} & \multicolumn{4}{c}{A4} \\
\cmidrule(lr){2-5} \cmidrule(lr){6-9}
Method & Frontal & Parietal & Occipital & Temporal & Frontal & Parietal & Occipital & Temporal \\
\midrule
GAN & 0.4457 & 0.5602 & 0.5200 & 0.5274 & 0.0922 & 0.3019 & 0.3769 & 0.1734 \\
VAE & 0.4463 & 0.5589 & 0.5115 & 0.5318 & 0.0147 & 0.0561 & 0.0319 & 0.0880 \\
DDPM & 0.5352 & 0.6683 & 0.6290 & 0.6255 & 0.3917 & 0.5779 & 0.4723 & 0.4494 \\
AnoDiff & 0.5321 & 0.6598 & 0.6225 & 0.6100 & 0.2905 & 0.5051 & 0.3964 & 0.3489 \\
Ours (D1) & \textbf{0.6561} & \textbf{0.7974} & \textbf{0.7372} & \textbf{0.7292} & \textbf{0.4382} & \underline{0.5842} & \textbf{0.4802} & \underline{0.4787} \\
Ours (D2) & \underline{0.6536} & \underline{0.7942} & \underline{0.7344} & \underline{0.7270} & \underline{0.4188} & \textbf{0.5942} & \underline{0.4799} & \textbf{0.4838} \\
\bottomrule
\end{tabular}
\end{table}

\begin{figure}
    \centering
    \includegraphics[width=1\linewidth]{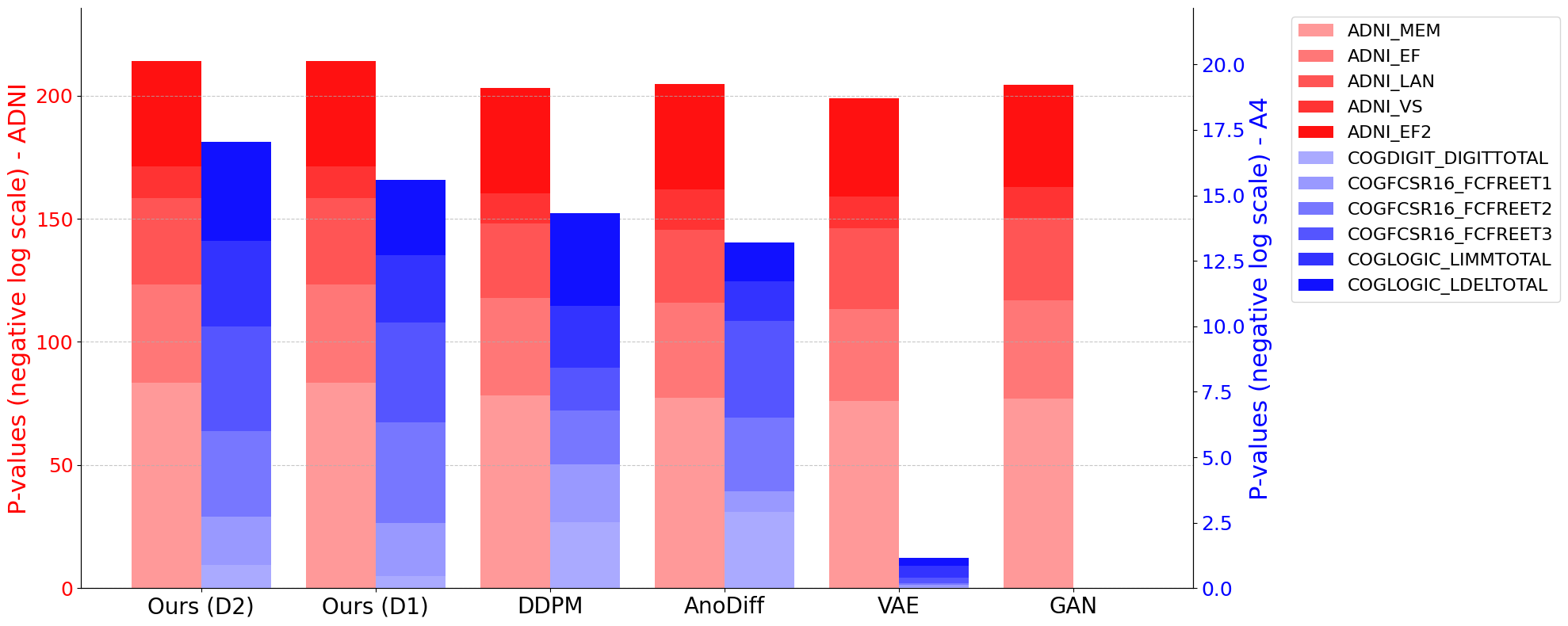}
    \caption{The p-value comparison on A4 and ADNI.}
    \label{fig:pvalue}
\end{figure}

We also quantitatively demonstrate the superior alignment of our derived anomaly score with SUVRs using Pearson correlation. The SUVRs are obtained by averaging the cortical SUVRs over the brain area for each subject. As shown in Table~\ref{tab:method_corr}, our methods achieve the highest correlation with SUVRs on both the ADNI and A4 datasets. In contrast, GAN and VAE baselines suffer from severe overfitting on the ADNI dataset, resulting in very low correlation values on the A4 cohort. As expected, the DDPM-based models outperform GAN and VAE. However, AnoDiff \cite{wolleb2022diffusion} performs comparably to vanilla DDPM on ADNI cohort but significantly worse on the preclinical A4 cohort. This is likely due to AnoDiff’s high sensitivity to the noising and denoising steps, leading to unstable performance across subjects with varying anomaly levels.

We evaluate the classification performance of our methods using the negative logarithm of the p-value for better visualization, as shown in Figure~\ref{fig:pvalue}. The ADNI dataset is used for training the CNN classifier, while the A4 dataset serves as the test set. Our results indicate that both VAE and GAN tend to overfit the ADNI training set, leading to poor classification performance in the preclinical A4 cohort. In contrast, DDPM and AnoDiff perform similarly on ADNI, with AnoDiff showing slightly worse performance on A4. Notably, our methods (D1 and D2) outperform all baselines on both datasets, highlighting their superiority in identifying preclinical subjects and their potential to aid in the early prevention of Alzheimer's disease progression.

\section{Conclusion}

In this paper, we propose a 3D latent bilateral-guided diffusion model, incorporating structure and intensity guidance, for tau PET anomaly localization. Our model employs deterministic sampling methods derived from approximations to stochastic differential equations and probability flow ODEs. Experimental results demonstrate superior performance on both clinical and preclinical datasets, outperforming baselines in localization and classification tasks. Moreover, our approach exhibits robustness against out-of-distribution data and contributes to early-stage cognitive impairment staging.

\begin{credits}
\subsubsection{\ackname}
This work was supported by the National Institute of Health (NIH) under grants RF1AG077578, RF1AG064584, R01EB022744, R21AG064776,\\ R01AG062007, U19AG078109, and P30AG066530. 
\end{credits}

\bibliographystyle{splncs04}
\bibliography{reference}

@inproceedings{tomasi1998bilateral,
  title={Bilateral filtering for gray and color images},
  author={Tomasi, Carlo and Manduchi, Roberto},
  booktitle={Sixth international conference on computer vision (IEEE Cat. No. 98CH36271)},
  pages={839--846},
  year={1998},
  organization={IEEE}
}

@article{choi2020development,
  title={Development and validation of language and visuospatial composite scores in ADNI},
  author={Choi, Seo-Eun and Mukherjee, Shubhabrata and Gibbons, Laura E and Sanders, R Elizabeth and Jones, Richard N and Tommet, Douglas and Mez, Jesse and Trittschuh, Emily H and Saykin, Andrew and Lamar, Melissa and others},
  journal={Alzheimer's \& Dementia: Translational Research \& Clinical Interventions},
  volume={6},
  number={1},
  pages={e12072},
  year={2020},
  publisher={Wiley Online Library}
}

@article{epstein2023diffusion,
  title={Diffusion self-guidance for controllable image generation},
  author={Epstein, Dave and Jabri, Allan and Poole, Ben and Efros, Alexei and Holynski, Aleksander},
  journal={Advances in Neural Information Processing Systems},
  volume={36},
  pages={16222--16239},
  year={2023}
}

@article{weigand2022s,
  title={What’s the cut-point?: a systematic investigation of tau PET thresholding methods},
  author={Weigand, Alexandra J and Maass, Anne and Eglit, Graham L and Bondi, Mark W},
  journal={Alzheimer's research \& therapy},
  volume={14},
  number={1},
  pages={49},
  year={2022},
  publisher={Springer}
}

@article{akamatsu2019voxel,
  title={Voxel-based statistical analysis and quantification of amyloid PET in the Japanese Alzheimer’s disease neuroimaging initiative (J-ADNI) multi-center study},
  author={Akamatsu, Go and Ikari, Yasuhiko and Ohnishi, Akihito and Matsumoto, Keiichi and Nishida, Hiroyuki and Yamamoto, Yasuji and Senda, Michio and Japanese Alzheimer’s Disease Neuroimaging Initiative},
  journal={EJNMMI research},
  volume={9},
  pages={1--9},
  year={2019},
  publisher={Springer}
}

@article{Mueller05,
title = "The Alzheimer's disease neuroimaging initiative",
author = "Mueller, {Susanne G.} and Weiner, {Michael W.} and Thal, {Leon J.} and Petersen, {Ronald C.} and Clifford Jack and William Jagust and Trojanowski, {John Q.} and Toga, {Arthur W.} and Laurel Beckett",
year = "2005",
month = {11},
volume = "15",
pages = "869--877",
journal = "Neuroimaging Clinics of North America",
number = "4"
}

@article{Sperling20,
    author = {Sperling, Reisa A. and Donohue, Michael C. and Raman, Rema and Sun, Chung-Kai and Yaari, Roy and Holdridge, Karen and Siemers, Eric and Johnson, Keith A. and Aisen, Paul S. and for the A4 Study Team},
    title = "{Association of Factors With Elevated Amyloid Burden in Clinically Normal Older Individuals}",
    journal = {JAMA Neurology},
    volume = {77},
    number = {6},
    pages = {735-745},
    year = {2020},
    month = {06}
}

@article{nelson2012correlation,
  title={Correlation of Alzheimer disease neuropathologic changes with cognitive status: a review of the literature},
  author={Nelson, Peter T and Alafuzoff, Irina and Bigio, Eileen H and Bouras, Constantin and Braak, Heiko and Cairns, Nigel J and Castellani, Rudolph J and Crain, Barbara J and Davies, Peter and Tredici, Kelly Del and others},
  journal={Journal of Neuropathology \& Experimental Neurology},
  volume={71},
  number={5},
  pages={362--381},
  year={2012},
  publisher={American Association of Neuropathologists, Inc.}
}

@article{SCHOLL2016971,
title = {{PET} Imaging of Tau Deposition in the Aging Human Brain},
journal = {Neuron},
volume = {89},
number = {5},
pages = {971-982},
year = {2016},
author = {Michael Schöll and Samuel N. Lockhart and Daniel R. Schonhaut and James P. O’Neil and Mustafa Janabi and Rik Ossenkoppele and Suzanne L. Baker and Jacob W. Vogel and Jamie Faria and Henry D. Schwimmer and Gil D. Rabinovici and William J. Jagust}
}

@inproceedings{yang2023learning,
  title={Learning with Synthesized Data for Generalizable Lesion Detection in Real PET Images},
  author={Yang, Xinyi and Chin, Bennett and Silosky, Michael and Litwiller, Daniel and Ghosh, Debashis and Xing, Fuyong},
  booktitle={International Conference on Medical Image Computing and Computer-Assisted Intervention},
  pages={116--126},
  year={2023},
  organization={Springer}
}

@inproceedings{siddiquee2019learning,
  title={Learning fixed points in generative adversarial networks: From image-to-image translation to disease detection and localization},
  author={Siddiquee, Md Mahfuzur Rahman and Zhou, Zongwei and Tajbakhsh, Nima and Feng, Ruibin and Gotway, Michael B and Bengio, Yoshua and Liang, Jianming},
  booktitle={Proceedings of the IEEE/CVF international conference on computer vision},
  pages={191--200},
  year={2019}
}

@article{salimans2022progressive,
  title={Progressive distillation for fast sampling of diffusion models},
  author={Salimans, Tim and Ho, Jonathan},
  journal={arXiv preprint arXiv:2202.00512},
  year={2022}
}

@article{jaeger2018digit,
  title={Digit symbol substitution test: the case for sensitivity over specificity in neuropsychological testing},
  author={Jaeger, Judith},
  journal={Journal of clinical psychopharmacology},
  volume={38},
  number={5},
  pages={513},
  year={2018},
  publisher={Wolters Kluwer Health}
}

@misc{psychometric2019,
  author = {{Knight Alzheimer's Disease Research Center}},
  title = {Psychometric Codebook},
  year = {2019},
  howpublished = {\url{https://knightadrc.wustl.edu/wp-content/uploads/2021/07/Psychometric-Codebook-7-22-19.pdf}},
  note = {Accessed: 2024-02-16}
}

@article{klein2009elastix,
  title={Elastix: a toolbox for intensity-based medical image registration},
  author={Klein, Stefan and Staring, Marius and Murphy, Keelin and Viergever, Max A and Pluim, Josien PW},
  journal={IEEE transactions on medical imaging},
  volume={29},
  number={1},
  pages={196--205},
  year={2009},
  publisher={IEEE}
}

@inproceedings{behrendt2024patched,
  title={Patched diffusion models for unsupervised anomaly detection in brain MRI},
  author={Behrendt, Finn and Bhattacharya, Debayan and Kr{\"u}ger, Julia and Opfer, Roland and Schlaefer, Alexander},
  booktitle={Medical Imaging with Deep Learning},
  pages={1019--1032},
  year={2024},
  organization={PMLR}
}

@article{kascenas2023role,
  title={The role of noise in denoising models for anomaly detection in medical images},
  author={Kascenas, Antanas and Sanchez, Pedro and Schrempf, Patrick and Wang, Chaoyang and Clackett, William and Mikhael, Shadia S and Voisey, Jeremy P and Goatman, Keith and Weir, Alexander and Pugeault, Nicolas and others},
  journal={Medical Image Analysis},
  volume={90},
  pages={102963},
  year={2023},
  publisher={Elsevier}
}

@inproceedings{wolleb2022diffusion,
  title={Diffusion models for medical anomaly detection},
  author={Wolleb, Julia and Bieder, Florentin and Sandk{\"u}hler, Robin and Cattin, Philippe C},
  booktitle={International Conference on Medical image computing and computer-assisted intervention},
  pages={35--45},
  year={2022},
  organization={Springer}
}

@inproceedings{rombach2022high,
  title={High-resolution image synthesis with latent diffusion models},
  author={Rombach, Robin and Blattmann, Andreas and Lorenz, Dominik and Esser, Patrick and Ommer, Bj{\"o}rn},
  booktitle={Proceedings of the IEEE/CVF conference on computer vision and pattern recognition},
  pages={10684--10695},
  year={2022}
}

@article{ho2020denoising,
  title={Denoising diffusion probabilistic models},
  author={Ho, Jonathan and Jain, Ajay and Abbeel, Pieter},
  journal={Advances in neural information processing systems},
  volume={33},
  pages={6840--6851},
  year={2020}
}

@article{song2020score,
  title={Score-based generative modeling through stochastic differential equations},
  author={Song, Yang and Sohl-Dickstein, Jascha and Kingma, Diederik P and Kumar, Abhishek and Ermon, Stefano and Poole, Ben},
  journal={arXiv preprint arXiv:2011.13456},
  year={2020}
}

@article{hassanaly2024evaluation,
  title={Evaluation of pseudo-healthy image reconstruction for anomaly detection with deep generative models: Application to brain FDG PET},
  author={Hassanaly, Ravi and Brianceau, Camille and Solal, Ma{\"e}lys and Colliot, Olivier and Burgos, Ninon},
  journal={arXiv preprint arXiv:2401.16363},
  year={2024}
}

@article{choi2019deep,
  title={Deep learning only by normal brain PET identify unheralded brain anomalies},
  author={Choi, Hongyoon and Ha, Seunggyun and Kang, Hyejin and Lee, Hyekyoung and Lee, Dong Soo},
  journal={EBioMedicine},
  volume={43},
  pages={447--453},
  year={2019},
  publisher={Elsevier}
}

@inproceedings{xiong2023pet,
  title={PET-3DFlow: A Normalizing Flow Based Method for 3D PET Anomaly Detection},
  author={Xiong, Zhe and Ding, Qiaoqiao and Zhao, Yuzhong and Zhang, Xiaoqun},
  booktitle={International Workshop on Computational Mathematics Modeling in Cancer Analysis},
  pages={91--100},
  year={2023},
  organization={Springer}
}

@article{han2021madgan,
  title={MADGAN: Unsupervised medical anomaly detection GAN using multiple adjacent brain MRI slice reconstruction},
  author={Han, Changhee and Rundo, Leonardo and Murao, Kohei and Noguchi, Tomoyuki and Shimahara, Yuki and Milacski, Zolt{\'a}n {\'A}d{\'a}m and Koshino, Saori and Sala, Evis and Nakayama, Hideki and Satoh, Shin’ichi},
  journal={BMC bioinformatics},
  volume={22},
  number={2},
  pages={1--20},
  year={2021},
  publisher={BioMed Central}
}

@article{baydargil2021anomaly,
  title={Anomaly analysis of Alzheimer’s disease in PET images using an unsupervised adversarial deep learning model},
  author={Baydargil, Husnu Baris and Park, Jang-Sik and Kang, Do-Young},
  journal={Applied Sciences},
  volume={11},
  number={5},
  pages={2187},
  year={2021},
  publisher={MDPI}
}
\end{document}